\documentclass[preprint,3p,12pt]{elsarticle}
\usepackage{lineno,hyperref}

\usepackage{amsmath}
\usepackage{verbatim} 
\usepackage{epsfig}
\usepackage{amssymb} 
\usepackage{mathtools}
\usepackage{lineno}
\usepackage{tikz}        
\usepackage{etoolbox}
\usepackage{ulem}         
\usepackage{textcomp}
\usepackage{nicematrix}
\usepackage{silence}
\usepackage{hyperref}
\usepackage{float}
\usepackage{subcaption}
\usepackage{graphicx}
\usepackage{placeins}

\journal{Physical Review E}
\begin{document}

\newcommand*{\hwplotB}{\raisebox{3pt}{\tikz{\draw[red,dashed,line 
width=3.2pt](0,0) -- 
(5mm,0);}}}

\newrobustcmd*{\mydiamond}[1]{\tikz{\filldraw[black,fill=#1] (0.05,0) -- 
(0.2cm,0.2cm) -- (0.35cm,0) -- (0.2cm,-0.2cm) -- (0.05,0);}}

\newrobustcmd*{\mytriangleright}[1]{\tikz{\filldraw[black,fill=#1] (0,0.2cm) -- 
(0.3cm,0) -- (0,-0.2cm) -- (0,0.2cm);}}

\newrobustcmd*{\mytriangleup}[1]{\tikz{\filldraw[black,fill=#1] (0,0.3cm) 
-- (0.2cm,0) -- (-0.2cm,0) -- (0,0.3cm);}}

\newrobustcmd*{\mytriangleleft}[1]{\tikz{\filldraw[black,fill=#1] (0,0.2cm) -- 
(-0.3cm,0) -- (0,-0.2cm) -- (0,0.2cm);}}
\definecolor{Blue}{cmyk}{1.,1.,0,0} 

\begin{frontmatter}

\title{The importance of the incubation time distribution in compartmental epidemiological models}

\address[IB]{Instituto Balseiro, Universidad Nacional de Cuyo}
\address[CONICET]{Consejo Nacional de Investigaciones Científicas y Técnicas (CONICET), Bariloche, Argentina}
\address[CNEA]{Gerencia de Física, Centro Atómico Bariloche, Comisión Nacional de Energía Atómica }

\author[IB,CONICET]{E.A.~Rozan}
\author[IB,CONICET,CNEA]{M.N.~Kuperman}
\author[CONICET,CNEA]{S.~Bouzat}

\begin{abstract}
\nolinebreak[0] This study investigates the utilization of various mathematical models for comprehending and managing outbreaks of infectious diseases, with a specific focus on how different distributions of incubation times influence predictions regarding epidemics. Two methodologies are examined: a compartmental SEnIR ODE model, which represents an enhanced version of the mean-field SEIR model, and a stochastic agent-based complex network model. Our findings demonstrate that the selection of diverse incubation time distributions can result in noteworthy discrepancies in critical epidemic forecasts, highlighting the crucial role of precise modeling in shaping effective public health interventions. The research underscores the necessity of integrating authentic distribution patterns into epidemic modeling to increase its reliability and applicability.

\end{abstract}


\end{frontmatter}


\setlength{\parskip}{12pt}
\section{Introduction} \label{section:intro}

The utilization of mathematical models in the analysis of outbreaks of infectious diseases has transformed our comprehension of disease propagation, and the implementation of more sophisticated models has emerged as an indispensable tool for enhancing preparedness and response within the realm of public health~\cite{keeling2008}. These models integrate various factors to predict an epidemic's duration and magnitude and offer valuable insights that guide decision-makers in taking well-informed choices. For instance, understanding the potential timing of infection peaks aids in the optimization of medical resource allocation, the securing of essential equipment and the establishment of supplementary medical facilities are essential in effectively treating patients during the critical phases of an outbreak.

Accurate foresight concerning mortality and infection rates facilitates pragmatic healthcare administration and the establishment of support systems for affected communities. Social, economic, and psychological assistance can be extended to families going through difficult situations. This empathetic approach is vital in maintaining social cohesion and mental well-being during periods of crisis~\cite{social_behavioural_science, Kitamura2022, impact_public_health}. Furthermore, estimating the potential size of an epidemic guides policy-makers in implementing appropriate interventions. Governments can discern the optimal timing and scope of measures such as quarantines, travel restrictions, and vaccination campaigns.

Most of the currently used models rely on ordinal differential equations and are rooted in the classical SIR models~\cite{ross1,kerm1}, and have a crucial role in studying epidemic outbreaks. A common addition to these models is the inclusion of an “exposed” stage, following infection yet preceding the infectious phase, during which the disease is incubating and cannot yet be transmitted to another individual. These extended models are categorized as SEIR models, representing a significant advancement in capturing the complex dynamics of disease transmission.

While these models are simple and flexible, their mathematical formulation imply an exponential distribution for the time that each individual remains in the exposed and infectious stages~\cite{waiting_time}. However, several studies reveal that the incubation times of most infectious diseases follow Gamma-like distributions~\cite{distribution_1949,hong_kong, malaria_plasmodium, estimating_incubation_period}, including the recent COVID-19 that originated a global pandemic~\cite{early_transmission_dynamics, canadian_context, incubation_period_COVID19, bahia_blanca, meta_analysis}. Some of these studies also suggest statistical methods to obtain the distribution during the initial phase of the outbreak.

To overcome this difficulty, some researchers have opted to adapt the SEIR model and  introduce additional \textit{ad-hoc} stages, so that the total time spent in those stages presents a Gamma distribution~\cite{appropriate_models, SEnImR_plantas, lloyd1,lloyd2}. This extended model, incorporating $n$ exposed stages, has received the name SE$n$IR. However, to our knowledge, there has not been a deep numerical analysis focusing on the impact of these supplementary exposed stages on pivotal epidemic variables such as the height of the infection peak, the time for the peak, and the epidemic size. Measuring how these quantities change for different distributions of the incubation time can potentially make a substantial difference when aiding decision-makers and is crucial for epidemic preparedness.

In this paper, we explore how the predictions can diverge from one another when considering different distributions for incubation times and how these biased predictions can affect their usefulness to public health agencies. 

Specifically, we will investigate how agent-based models in complex networks handle this issue, offering flexibility in selecting time distributions. We show that in a fully connected network, the results coincide with the SEIR ODE-model when the distribution is exponential, and notably differs when a non-exponential distribution is considered for the incubation time. We will also show results for agents in a scale-free network to contrast the results derived from the present and traditional mean field-models.

In Section~\ref{section:compartmental}, we employ the mean field approach to address this matter. In Section \ref{section:network}, we transition to the agent-based complex network approach. The results are discussed in Section~\ref{section:discussion}, while the conclusions are given in Section~\ref{section:conclusion}.

\section{Mean-field SE\textit{n}IR Model} \label{section:compartmental}
\subsection{Introduction to the SE\textit{n}IR Model}
The starting point for the present work is the classical SEIR model. This fundamental mean-field compartmental model divides individuals into four distinct states or compartments: S (susceptible to infection), E (exposed to the infection but not yet infectious), I (infected and capable of transmitting the infection), and R (recovered individuals, including both those who have recovered from the illness and those who have succumbed to it). The evolution of the infection is described by the following equations:
\begin{equation}
\begin{NiceMatrix}[l]
\displaystyle\frac{\text d S}{\text d t} =&
\displaystyle -rSI\\[0.5cm]
\displaystyle \frac{\text d E}{\text d t} =&
\displaystyle rSI - \frac{E}{T_\text{inc}}\\[0.5cm]
\displaystyle \frac{\text d I}{\text d t} =&
\displaystyle  \frac{E}{T_\text{inc}} - \frac{I}{T_\text{inf}}\\[0.5cm]
\displaystyle\frac{\text d R}{\text d t} =&
\displaystyle  \frac{I}{T_\text{inf}}\end{NiceMatrix}
\label{SEIR}
\end{equation}
\noindent where $S+E+I+R=1$, as each variable represents the fraction of the individuals in the respective compartment. Here, $r$ is the mean effective infection rate, $T_\text{inc}$ is the mean incubation time and $T_\text{inf}$ is the mean infection time. 

As stated in the previous section, Eqs.~(\ref{SEIR})  are associated with a model that assumes an exponential distribution for the duration of the incubation period, $\tau\sim\exp(T_\mathrm{inc})$. However, it has been suggested that a Gamma-type distributions fit field observations much better~\cite{distribution_1949,hong_kong, malaria_plasmodium, estimating_incubation_period,early_transmission_dynamics, canadian_context, incubation_period_COVID19, bahia_blanca, meta_analysis}. A straightforward way to obtain a Gamma distribution for the time of stay in a given stage is to divide the corresponding stage into $n$ sub-stages, each one with an exponentially distributed waiting time. 

To show the validity of this method we can consider the moment generating function (MGF) of the probability distribution of a real-valued random variable~\cite{Bulmer}. If we define 
\begin{equation}
 S=\sum^{n}_{i=1} X_i    
\end{equation}
as the sum of independent and identically distributed  random variables $X_i$, each with MGF $M_X(t)$, then the MGF of $S$ is
\begin{equation}
M_S(t)=\left[M_X(t)\right]^n
\end{equation}
If we consider that the variables $X_i$ are exponentially distributed, $X_i\sim\text{Exp}(\lambda)$, with an MGF 
\begin{equation}
M_X(t)=\frac{\lambda}{\lambda-t}
\end{equation}
then we get the following MGF for the variable $S$:
\begin{equation}
M_S(t)=\left[M_X(t)\right]^n=\left [\frac{\lambda}{\lambda-t}\right]^{n}
\end{equation}
which is the MGF for a Gamma distribution $\Gamma(n,\lambda)$. This shows that $S$ has a Gamma distribution~\cite{Bulmer}.

In the context of this model, we will divide the exposed compartment $E$ into $n$ sub-compartments $E_i$, each one with a waiting time $\tau_i$ having an exponential distribution with mean time $T_\mathrm{inc}/n$. The total incubation time $\tau$ results from the sum of the partial time in each sub-stage. According to what we have shown above, it has the following Gamma distribution:
\begin{equation}
\tau_i\sim\mathrm{Exp}(T_\mathrm{inc}/n) \Longrightarrow \tau = \sum_{i=1}^n \tau_i  \sim \mathrm{\Gamma}(n,T_\mathrm{inc}/n)
\label{gamma_distribution}
\end{equation}

\begin{figure}[ht]\centering
\begin{subfigure}{.49\textwidth} \centering
\includegraphics[width=\linewidth]{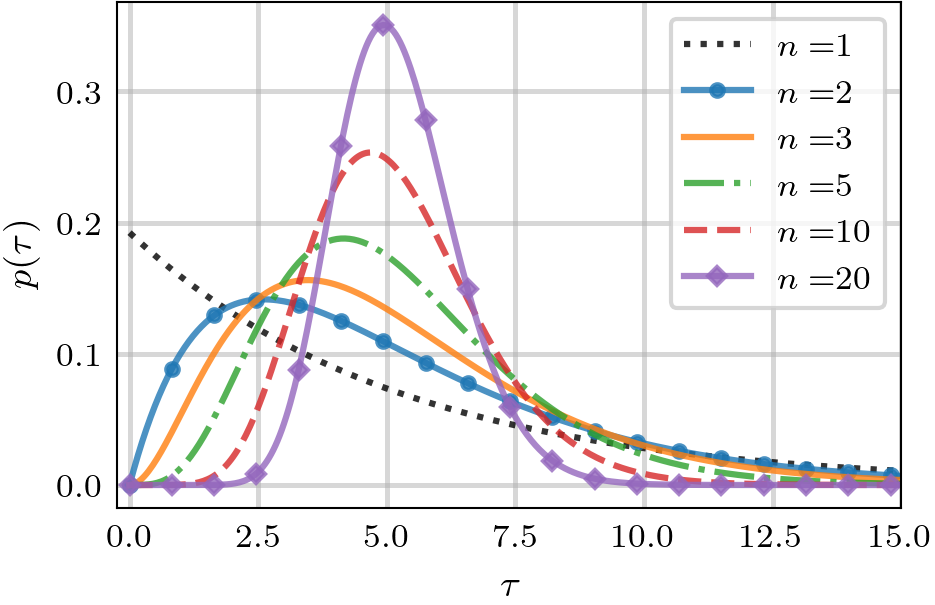}
\caption{}
\label{gamma_examples}
\end{subfigure}
\begin{subfigure}{.49\textwidth} \centering
\includegraphics[width=\linewidth]{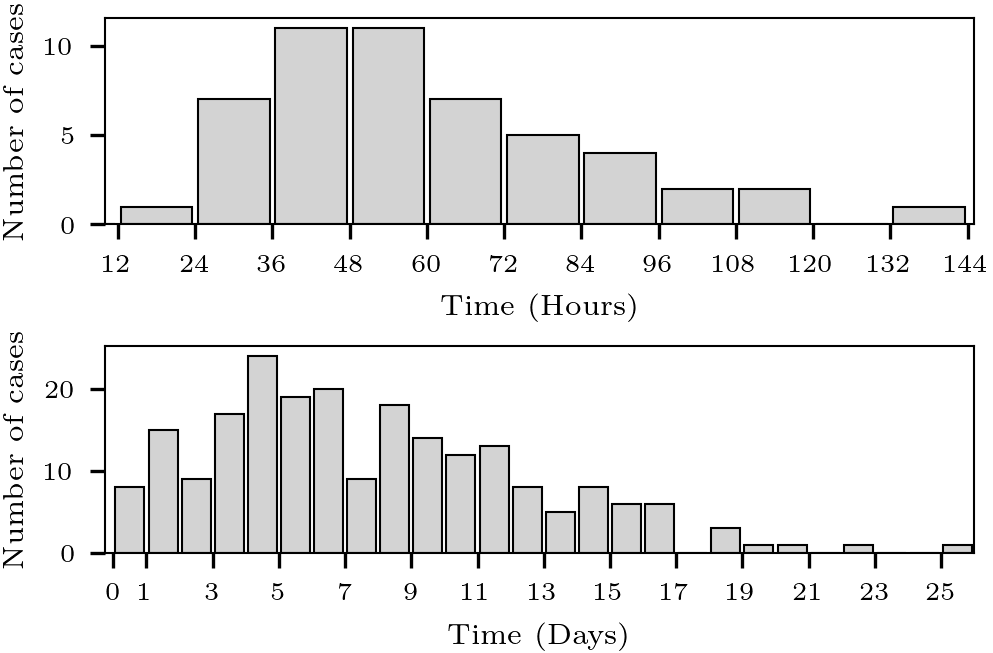}
\caption{}
\label{dist_experimentales}
\end{subfigure}
\caption{(a) Plot of the Gamma distribution $\mathrm{\Gamma}(n,T_\mathrm{inc}/n)$ for $T_\mathrm{inc}=5.2$ and several values of $n$. (b)~Experimental data of the incubation period in an epidemic of milk-borne streptococcal sore throat~\cite{distribution_1949} (top graph), and in the COVID-19 pandemic~\cite{meta_analysis} (bottom graph). }
\end{figure}

As the mean value of the Gamma distribution $\Gamma(n,\lambda)$ is $n\lambda$~\cite{Bulmer}, the distribution of the total incubation time $\tau$ shown in Eq.~(\ref{gamma_distribution}) has mean $T_\mathrm{inc}$ for any value of $n$. Examples of the probability distribution functions are plotted in Fig.~\ref{gamma_examples}. Note that for $n=1$ the distribution is exponential, as expected, and peaks at $\tau=0$, which differs from all field observations. However for higher $n$ the maximum of the distribution is attained at longer times. As $n$ increases, the distribution tends towards a normal distribution with lower variance, which is consistent with the central limit theorem. The selection of  a value for $n$ determines the distribution of incubation times, a parameter that can be promptly estimated during the initial phases of an outbreak (as shown in \cite{early_transmission_dynamics,incubation_period_COVID19}, which analyzed the incubation time distribution of COVID-19 in the early stages of 2020, for instance). 

Fig.~\ref{dist_experimentales} shows the distribution for the incubation times of streptococcal sore throat and COVID-19, as experimentally measured in~\cite{distribution_1949} and~\cite{meta_analysis}, respectively. Estimating the mean value in each case and then fitting the histograms to Gamma distributions yields a parameter of $n=2$ and $n=6$, respectively, reflecting the fact that different infectious diseases can exhibit distributions better suited to different values of $n$. Nonetheless, it is evident that the times are not exponentially distributed.

\FloatBarrier

When considering $n$ exposed sub-stages, the system of equations is as follows:
\begin{equation}
\begin{NiceMatrix}[l]
\displaystyle \frac{\text d S}{\text d t} =&
\displaystyle -rSI\\[0.5cm]
\displaystyle\frac{\text d E_1}{\text d t} =&
\displaystyle \color{white}-\color{black} rSI - \frac{E_1}{T_\text{inc}/n}\\[0.5cm]
\displaystyle\frac{\text d E_i}{\text d t} =&
\displaystyle \color{white}-\color{black} \frac{E_{i-1}}{T_\text{inc}/n}-  \frac{E_{i}}{T_\text{inc}/n}\qquad i=2,3,...,n\\[0.5cm]
\displaystyle\frac{\text d I}{\text d t} =&
\displaystyle  \phantom{-}\frac{E_n}{T_\text{inc}/n} - \frac{I}{T_\text{inf}}\\[0.5cm]
\displaystyle\frac{\text d R}{\text d t} =&
\displaystyle \color{white}-\color{black} \frac{I}{T_\text{inf}}\end{NiceMatrix}
\label{SEnIR}
\end{equation}

In the present work we will consider $T_\mathrm{inc}=5.2$ days and $T_\mathrm{inf}=14-T_\mathrm{inc}=8.8$ days, which are the estimated values of these parameters for the recent COVID-19 pandemic~\cite{early_transmission_dynamics}. This study however does not focus particularly on this case, as our goal is to study the effect that the distribution of incubation time has on the model in general.

We will not fix a value for $r$, since it will be varied to obtain specific values for the doubling time. In the midst of epidemic outbreaks, such as the recent COVID-19 pandemic, the doubling time emerges as a readily measurable parameter, serving as a key indicator of the outbreak's severity \cite{Kapitany, Carroll2020}. Additionally, it is an easily accessible parameter for the general public, often published through official reports. We recall that the doubling time represents the duration it takes for the total infected population $C(t) = I(t) + R(t)$ to double when it follows exponential growth, expressed by $C(t + D) = 2C(t)$. When a model is chosen to fit into any one specific scenario, the doubling time obtained should coincide with the experimentally measured one. Thus, to obtain the same doubling time $D$ for different values of $n$, we gradually adjust the infection rate $r$ until the doubling time fits the desired one. 

As the initial condition we consider $S(0)=0.99$, $I(0)=0.01$, and ${E_{i}(0)=R(0)=0}$. 

\subsection{Results}

\begin{figure}[ht]\centering
\begin{subfigure}{.325\textwidth} \centering
\includegraphics[width=\linewidth]{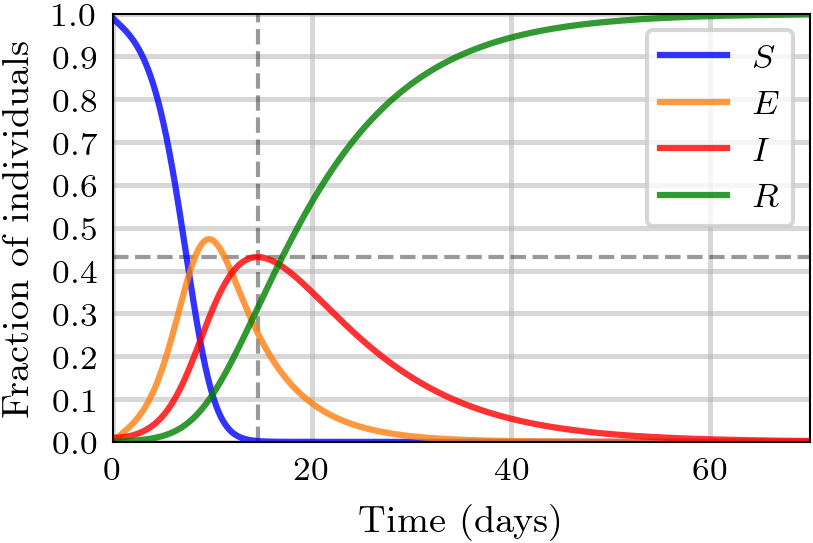}
\caption{$n=1$}
\label{SEIR_SE5IR_D2_a}
\end{subfigure}
\begin{subfigure}{.325\textwidth} \centering
\includegraphics[width=\linewidth]{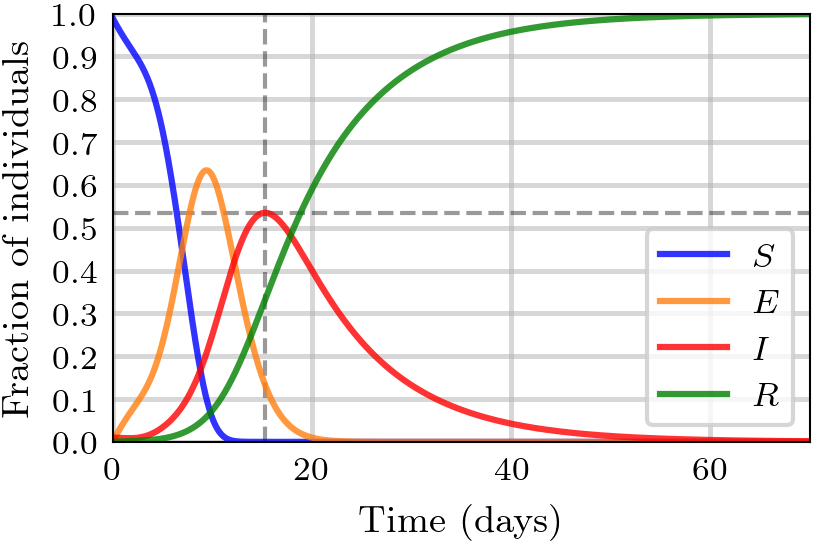}
\caption{$n=5$}
\label{SEIR_SE5IR_D2_b}
\end{subfigure}
\begin{subfigure}{.325\textwidth} \centering
\includegraphics[width=\linewidth]{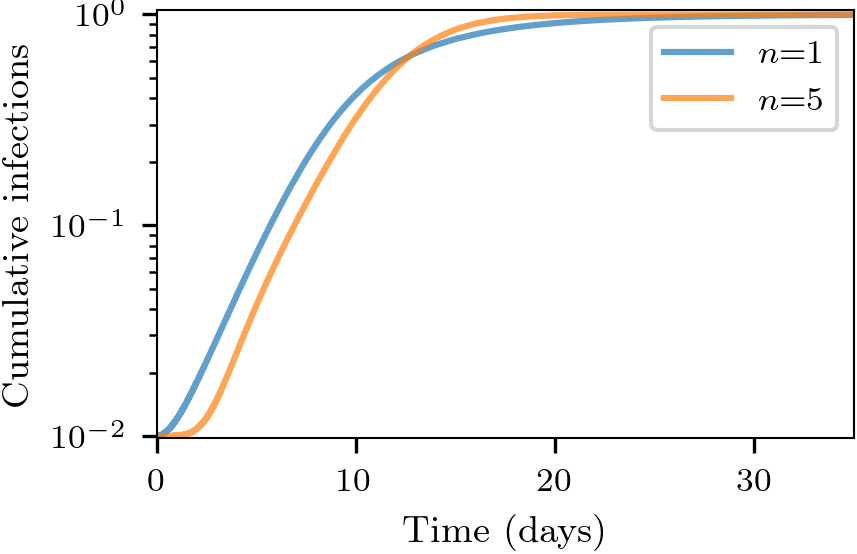}
\caption{$C(t)=I(t)+R(t)$}
\label{SEIR_SE5IR_D2_c}
\end{subfigure}
\caption{(a) \& (b) Evolution of the fraction of individuals in each compartment for the SE$n$IR model with doubling time $D=2$~days. The dashed lines show the time and height of the infection peak. (c) Cumulative fraction of infected individuals for the systems in (a) and (b), that grow parallel to one another in logarithmic scale.}
\label{SEIR_SE5IR_D2}
\end{figure}

First, we examine the temporal progression of the infection under two distinct scenarios: one involving an exponential distribution and the other a Gamma distribution for the duration of the Exposed stage. Figure~\ref{SEIR_SE5IR_D2} shows this evolution, displaying  the fraction of individuals in each compartment for $n=1$ and $n=5$, where the curve 
showing the exposed compartment corresponds to  $E(t)=\sum_{i=1}^nE_i(t)$ in all cases. The doubling time is $D=2$~days in both cases, which makes the cumulative fraction of infected individuals grow parallel to one another, as shown in Fig.~\ref{SEIR_SE5IR_D2_c}. Recall that we chose to maintain $D$ fixed as it is one of the indicators that is most straightforwardly measured. Therefore, to properly compare results from different $n$, the doubling time should be the same. The value of $r$ is chosen in each case to obtain the desired doubling time, $D=2$~days in the case of Figure~\ref{SEIR_SE5IR_D2}. 

Notably, the peak of $I(t)$ is 23.8\% higher for $n=5$ when compared to $n=1$. Thus, the prediction made by the most commonly used model with exponential distributions can be severely underestimating the infection peak. 

\begin{figure}[ht]\centering
 \includegraphics[width=0.6\linewidth]{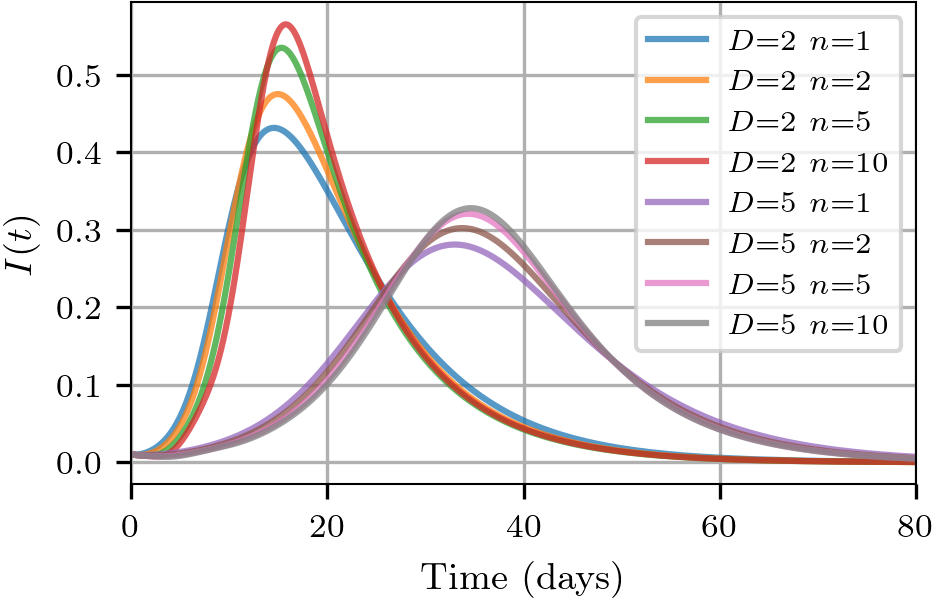}
 \caption{Infected curve as a function of time, $I(t)$, for $D=2$ and $D=5$, and $n=1, 2, 5$ and $10$. }
\label{many_I_Dfijo}
\end{figure}

While both figures show qualitatively the same results, the relevance of their differences will be revealed in the following plots. Fig.~\ref{many_I_Dfijo} compares the infection curves more directly, for different values of $n$, and for $D=2$ and $D=5$~days. Note that the curves with $D=2$ grow apart from one another as $n$ increases, more so than the curves with $D=5$~days do.

\begin{figure}[ht]\centering
 \begin{subfigure}{.325\textwidth} \centering
 \includegraphics[width=\linewidth]{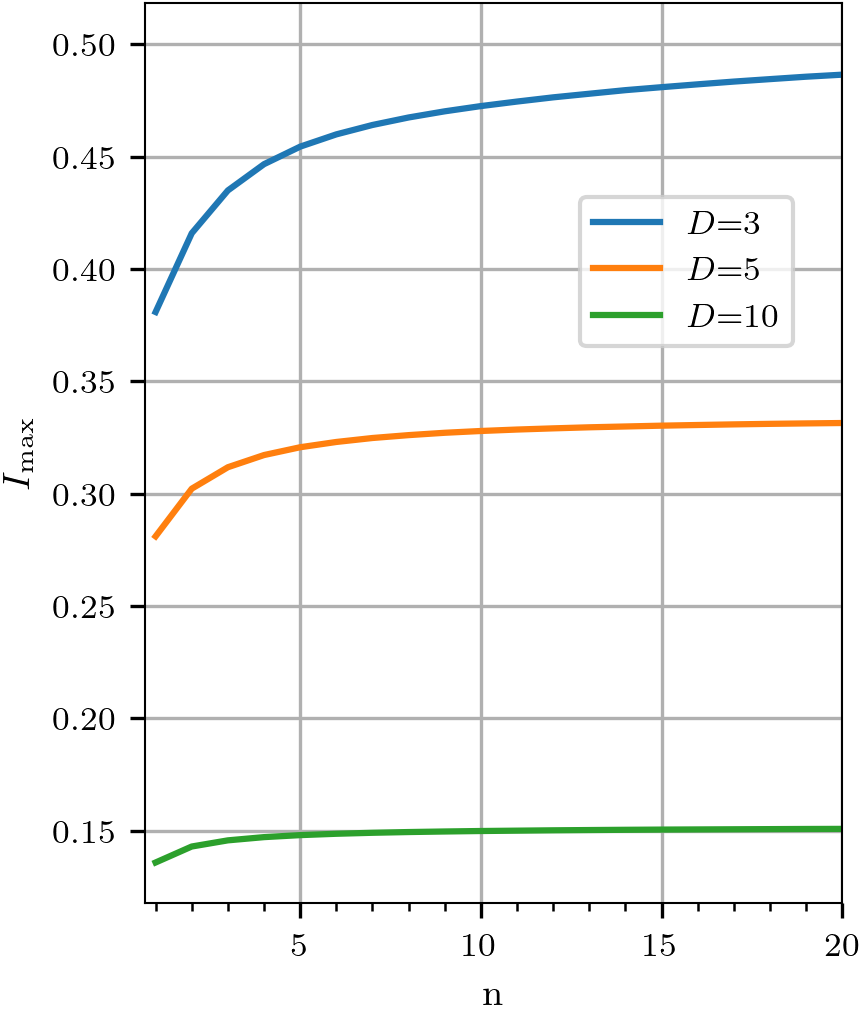}
 \caption{$I_\text{max}$}
 \end{subfigure}
  \begin{subfigure}{.325\textwidth} \centering
 \includegraphics[width=\linewidth]{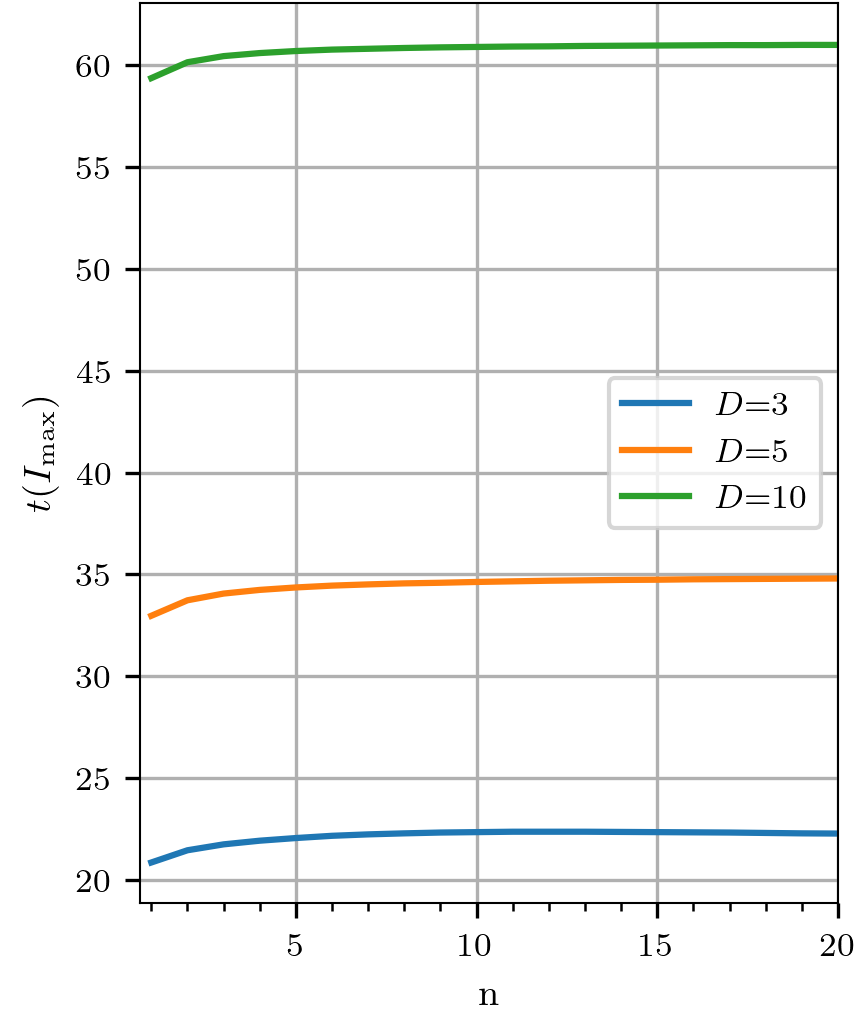}
 \caption{ $t(I_\text{max})$}
 \end{subfigure}
   \begin{subfigure}{.325\textwidth} \centering
 \includegraphics[width=\linewidth]{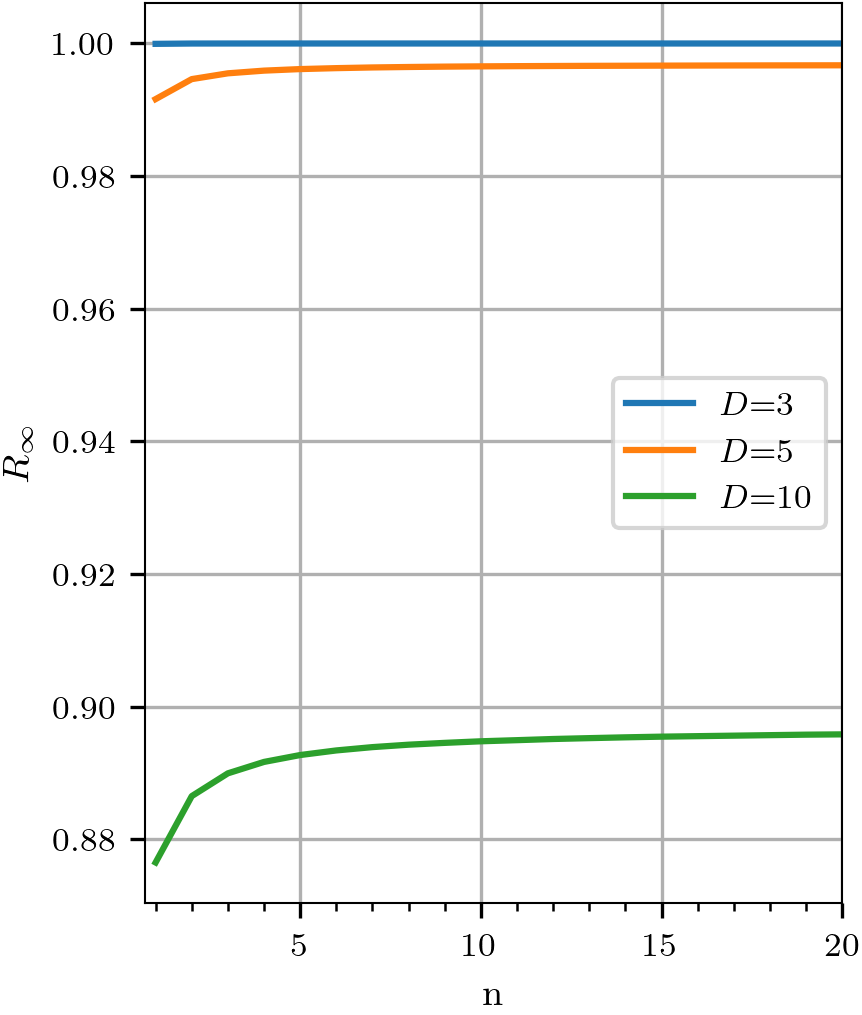}
 \caption{ $R_\infty$}
 \end{subfigure}
 \caption{Relevant quantities as a function of $n$, for fixed values of the doubling time $D$. }
\label{obs_vs_n}
\end{figure}

To delve deeper into what was observed in the previous figure, we analyze in detail the behavior of the temporal evolution infection curves, emphasizing the behavior of relevant quantities such as the height of the infection peak $I_\mathrm{max}$, the time at which it occurs $t(I_\mathrm{max})$, and also the epidemic size $R_\infty=R(t=\infty)$, each one as a function of $n$. The results are shown in Fig.~\ref{obs_vs_n}

\begin{figure}[h]\centering
 \begin{subfigure}{.325\textwidth} \centering
 \includegraphics[width=\linewidth]{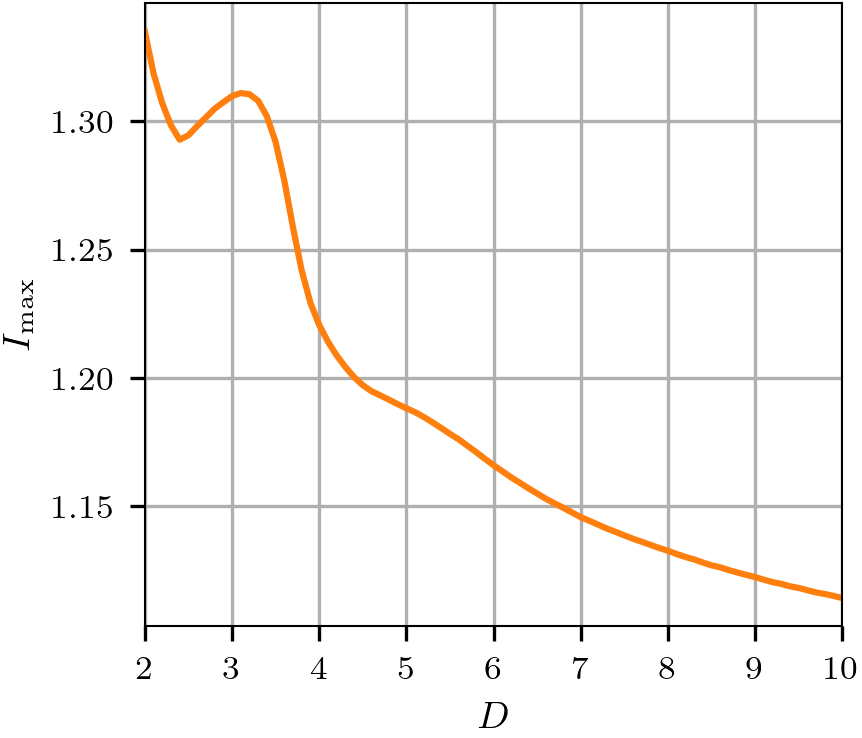}
 \caption{$I_\text{max}$}
 \end{subfigure}
  \begin{subfigure}{.325\textwidth} \centering
 \includegraphics[width=\linewidth]{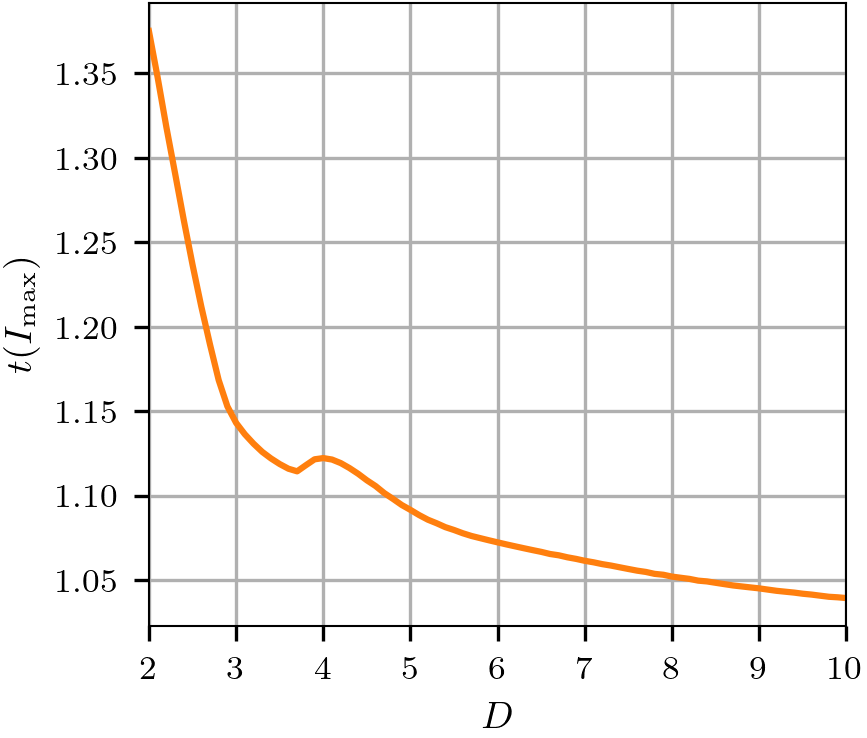}
 \caption{ $t(I_\text{max})$}
 \end{subfigure}
   \begin{subfigure}{.325\textwidth} \centering
 \includegraphics[width=\linewidth]{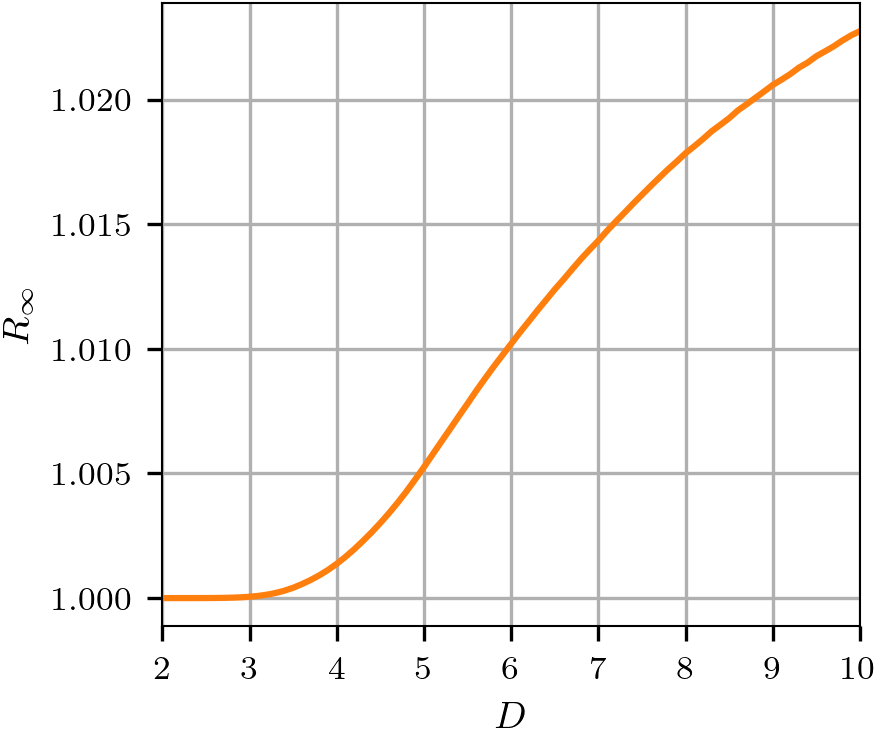}
 \caption{$R_\infty$}
 \end{subfigure}
 \caption{Ratio between the maximum and minimum value attained by each quantity for a given $D$.}
\label{cocientes_vs_D}
\end{figure}

Then we analyzed the behavior of each of the quantities displayed in Fig.~\ref{obs_vs_n} as a function of $n$, with $n$ running between 0 and 50 and a fixed value of $D$. To characterize the obtained results in a compact way we took the maximum and minimum values obtained for each quantity, and calculated the ratio between them as a function of $D$. The results are displayed in Fig.~\ref{cocientes_vs_D}, and show, for instance, that the height of the infection peak can differ by more than $30\%$ depending on the incubation time distribution chosen. The same is true for the time to reach the the peak. The epidemic size, on the contrary, does not change in a significant way.

Fig.~\ref{cocientes_vs_D} also shows that the differences in $I_\text{max}$ and $t(I_\text{max})$ are more pronounced when the doubling time is shorter. These are the most alarming cases, in which the infection curve exhibits a more rapid growth. Hence, accurate prediction of these quantities becomes especially crucial in such cases.

In this section we focused on the changes in the predictions of relevant quantities and how they diverge when considering different distributions of incubation times in mean-field models. We will next address the same question but in the context of agent-based models in complex networks. 

\FloatBarrier
\section{Stochastic agent-based complex network model}\label{section:network}

In a mean-field model, the variations in the time that the individuals spend in each epidemiological stage are evident as the transition rates between states mirror the statistical nature of waiting times within each stage. However, it's important to note that the distribution of these times always results in a Gamma distribution, and remains the same across all individuals. In contrast, in an agent-based SEIR model characterized by distinctly individualized agents, we have the capacity to explicitly define and customize the incubation time distribution. This enables us to investigate the same aspect as the one discussed in the preceding section but from a distinct angle. This also allows different time distributions among the individuals, should field reports indicate different distributions based in the age, gender, or other characteristics of the population, although this is out of the scope of this work.

The model we present here comprises $N$ nodes within a complex network, with each node representing an individual. Each one of them can be in one of four epidemic compartments (S, E, I, or R), and time advances discretely in steps. At each time-step, a susceptible individual can contract the infection from each of his (her) infectious neighbors with probability $r$, and transition to the exposed state. That means that an individual denoted as $j$ can contract the infection with probability  $1-(1-r)^{k^I_j}$, where $k^I_j$ is the total amount of infectious neighbors of the individual $j$. Subsequently, the individual remains exposed for a random duration generated from a specified probability distribution before becoming infectious. Likewise, the individual will recover after a random infection period, generated from a different probability distribution.

In this section, we present results obtained with $N=10^6$ agents, a time-step of $\Delta t = (24\ \mathrm{days})^{-1}$ (equivalent to one hour, making 24 time-steps equal one day), and a contagion rate of $r=\frac{0.5 \Delta t}{N}\simeq4.16\cdot 10^{-8}$. We opted to maintain $r$ constant due to the challenge of fitting the cumulative infection curve to achieve a desired doubling time, given the stochastic nature and duration of the simulations. The distribution for the infection time remains fixed at Exp$(T_\mathrm{inf})$ with $T_\mathrm{inf}=8.8$ days, while the distribution for the incubation time varies. However, its mean value remains fixed at $T_\mathrm{inc}=5.2$ days, consistent with Section~\ref{section:compartmental}.

To contrast the findings in this section with those in the preceding one, we begin by presenting results pertaining to fully connected networks, as they constitute the primary assumption of mean-field compartmental models like the one outlined by Eq. (1).

\begin{figure}[ht]\centering
 \begin{subfigure}{.49\textwidth} \centering
 \includegraphics[width=\linewidth]{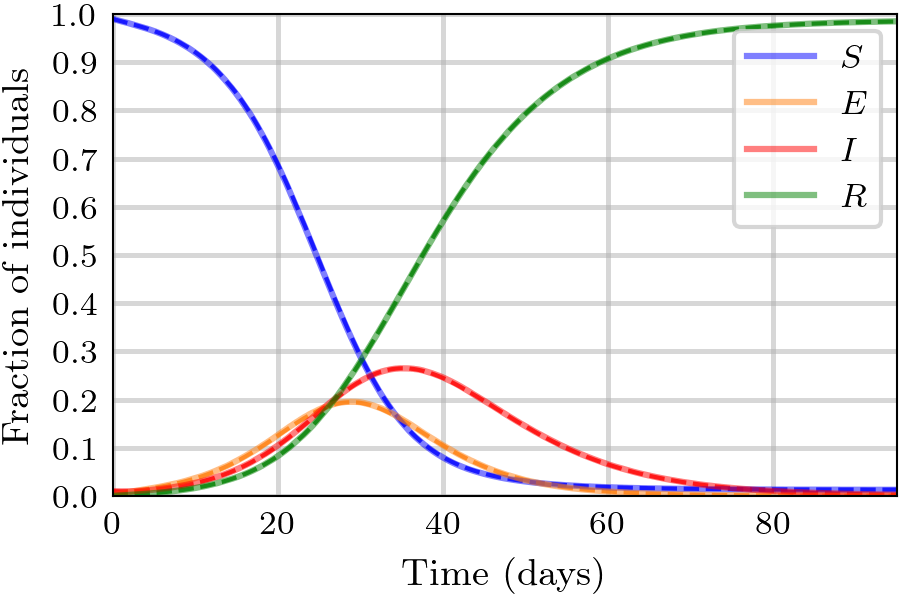}
 \caption{Exponential distribution for incubation times}
 \label{full_graph_exp}
 \end{subfigure}
  \begin{subfigure}{.49\textwidth} \centering
 \includegraphics[width=\linewidth]{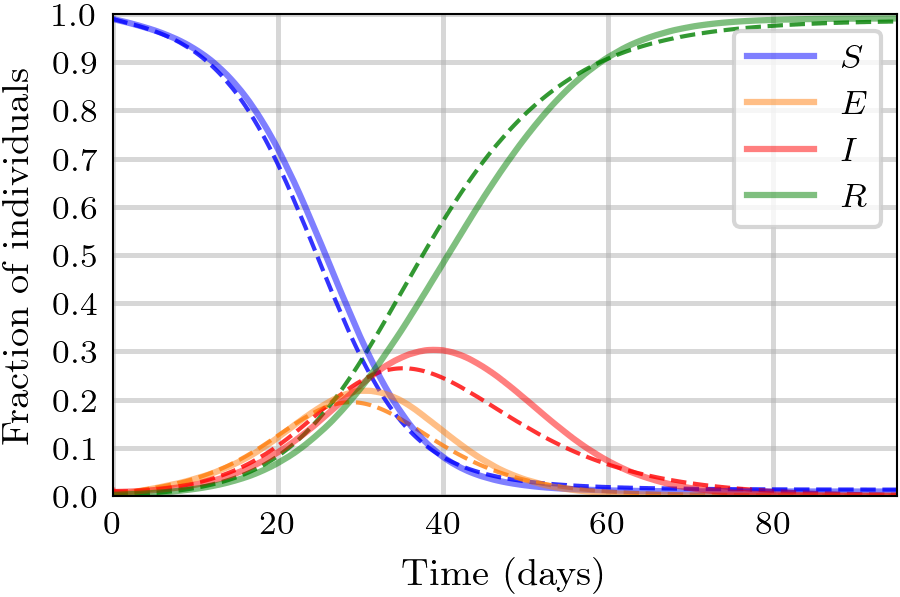}
 \caption{Uniform distribution for incubation times}
  \label{full_graph_unif}
 \end{subfigure}
 \caption{Evolution of the fraction of individuals in each compartment in fully connected networks for different incubation time distributions. Dashed lines correspond to the solution to the classical SEIR ODE model from Eq.~(\ref{SEIR}).}
\label{full_graph}
\end{figure}

Fig.~\ref{full_graph} shows the evolution of the fraction of individuals in each compartment as a function of time in solid lines, while dashed lines correspond to the solution to the most classical SEIR ODE model from Eq.~(\ref{SEIR}). In Fig.~\ref{full_graph_exp} the incubation times follow an exponential distribution, and we can see that the results using this distribution are practically indistinguishable from the mean-field model. On the other hand, in Fig.~\ref{full_graph_unif} the incubation time follows a Uniform($T_\text{inc}-\sqrt3,T_\text{inc}+\sqrt3$) distribution, which has the same mean and variance as the exponential one. In this case, the evolution of the system is clearly different. For instance, the $I_\mathrm{max}$ obtained for the exponential distribution is $0.2671$, while for the uniform distribution, it is $0.3031$, which is $13.5\%$ more. We can see that even for fully connected networks, a different time incubation distribution produces a notable difference in the prediction. 

\FloatBarrier

\begin{figure}[ht]\centering
 \begin{subfigure}{.49\textwidth} \centering
 \includegraphics[width=\linewidth]{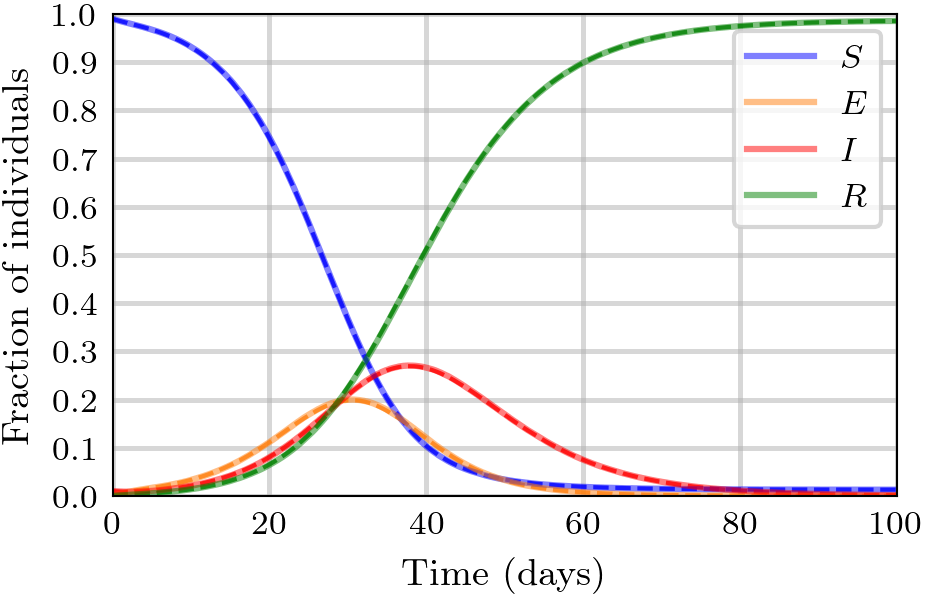}
 \caption{$n=2$}
 \end{subfigure}
  \begin{subfigure}{.49\textwidth} \centering
 \includegraphics[width=\linewidth]{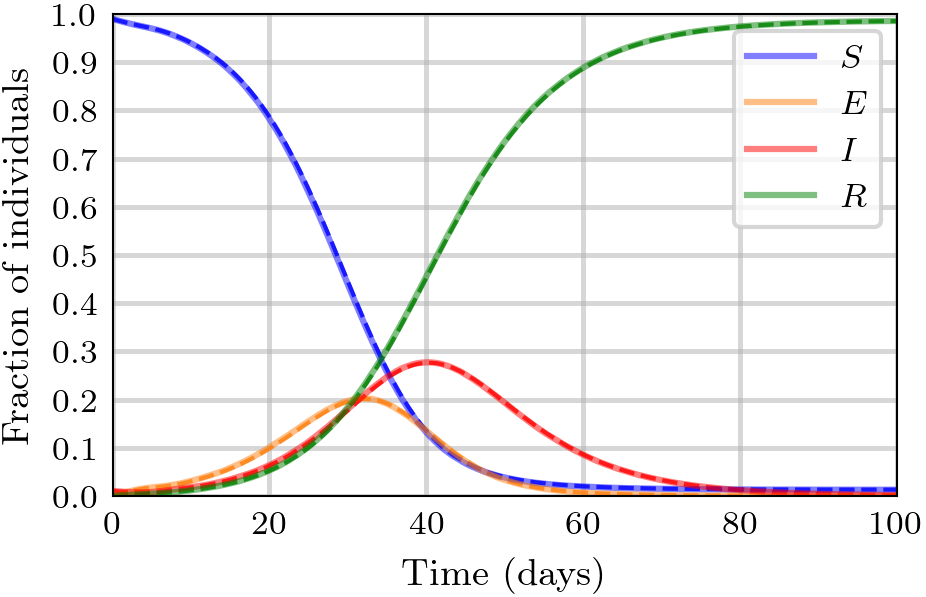}
 \caption{$n=10$}
 \end{subfigure}
 \caption{Evolution of the fraction of individuals in each compartment in fully connected networks, when the incubation time distributions follow a Gamma distribution $\Gamma(n,T_\mathrm{inc}/n)$. Dashed lines correspond to the solution of the SE$n$IR ODE model from Eq.~(\ref{SEnIR}) with the corresponding $n$.}
\label{full_graph_gamma}
\end{figure}

\begin{figure}[ht]\centering
 \begin{subfigure}{.325\textwidth} \centering
 \includegraphics[width=\linewidth]{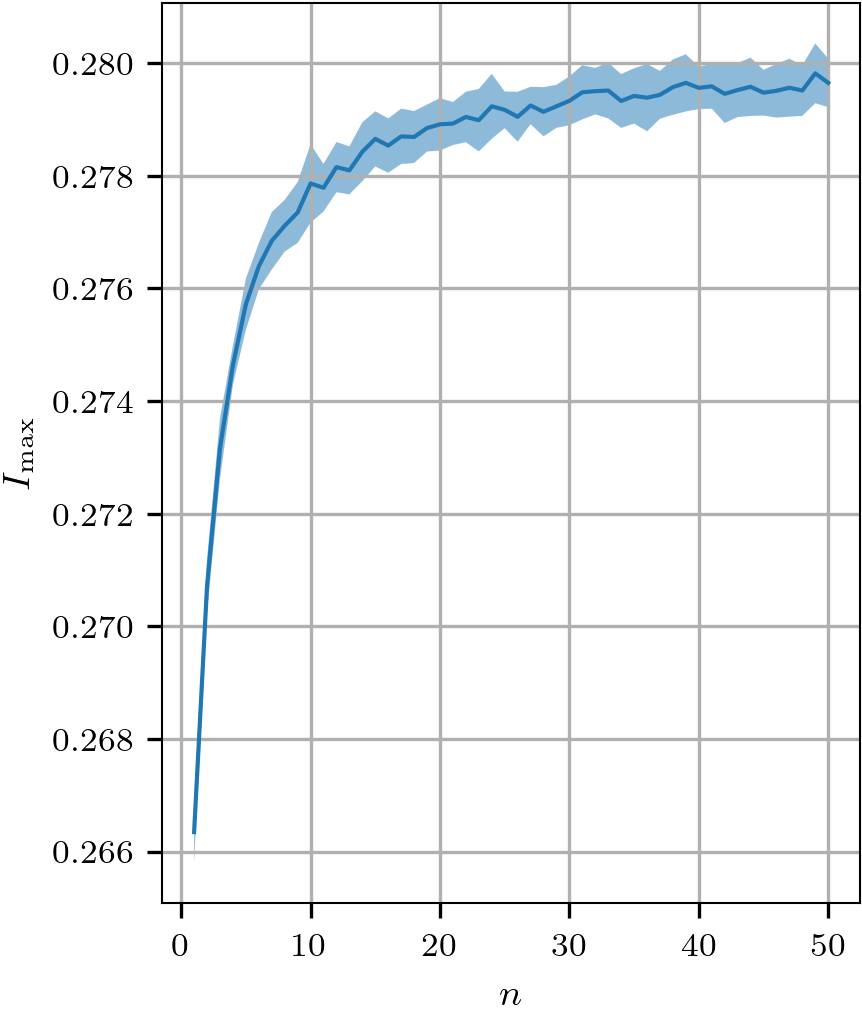}
 \caption{$I_\text{max}$}
 \end{subfigure}
  \begin{subfigure}{.325\textwidth} \centering
 \includegraphics[width=\linewidth]{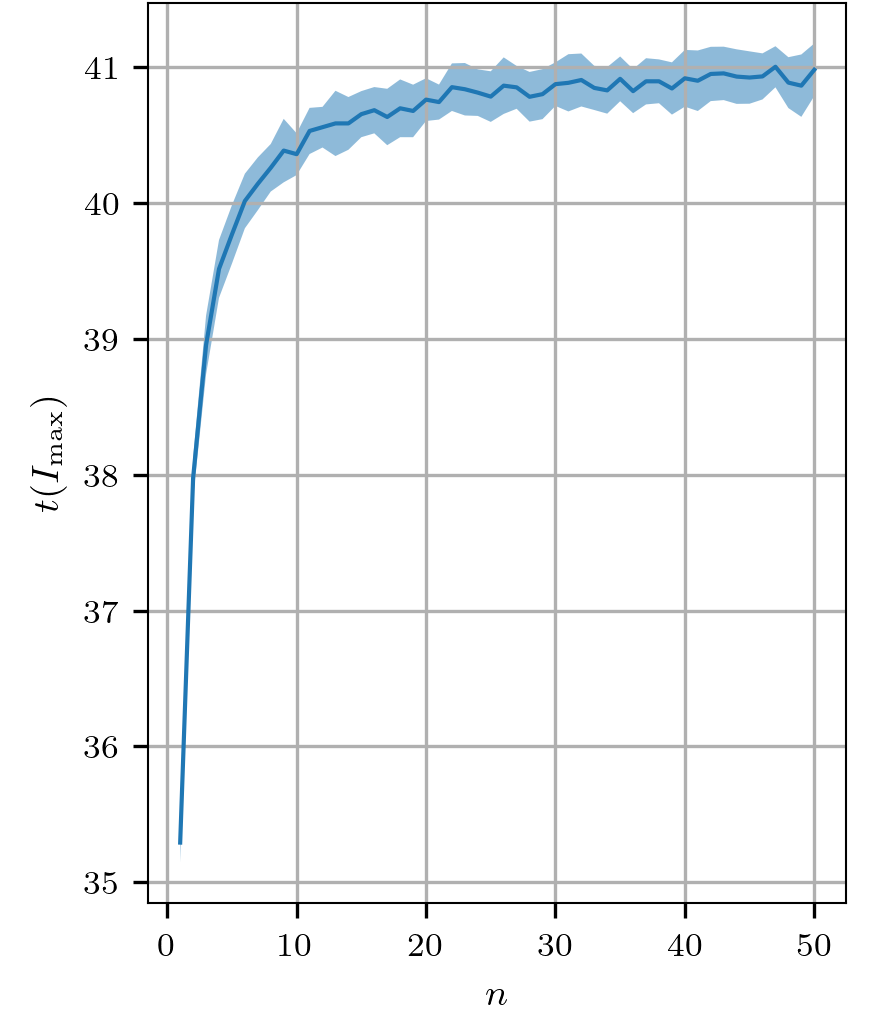}
 \caption{ $t(I_\text{max})$}
 \end{subfigure}
   \begin{subfigure}{.325\textwidth} \centering
 \includegraphics[width=\linewidth]{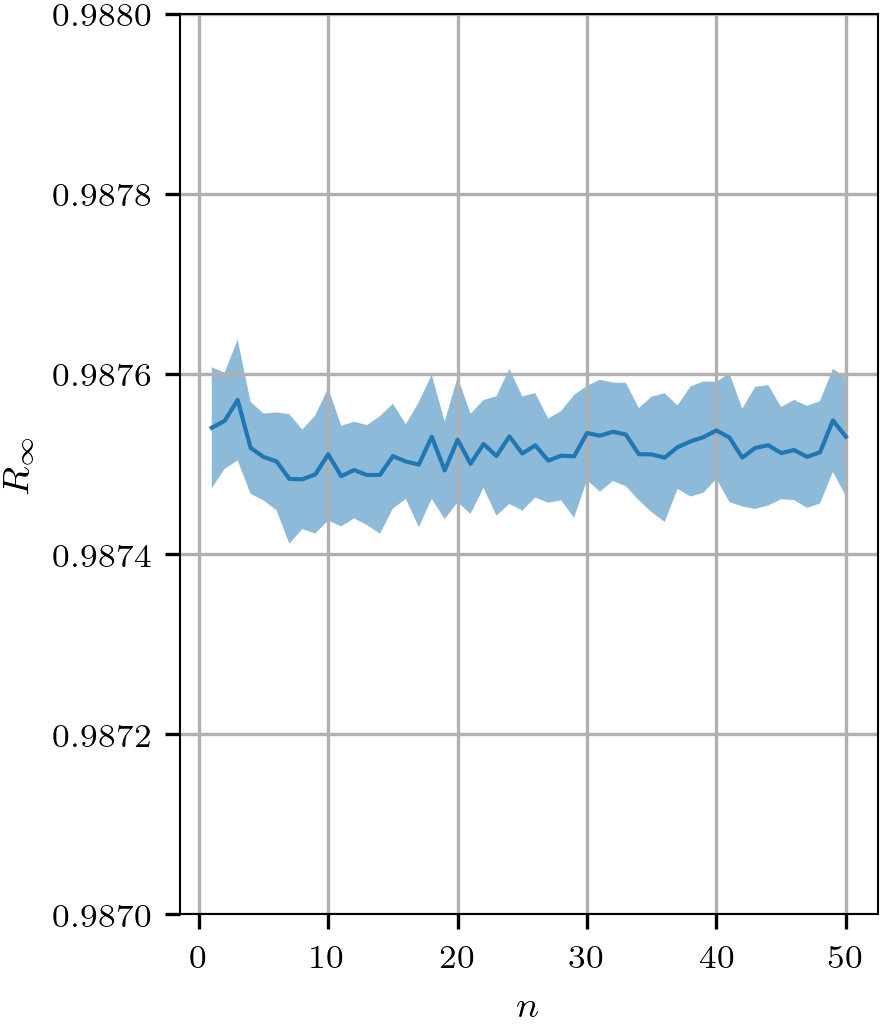}
 \caption{ $R_\infty$}
 \end{subfigure}
 \caption{$I_\text{max}$, $t(I_\text{max})$ and $R_\infty$ as a function of $n$ in fully connected networks.}
\label{full_graph_obs}
\end{figure}

When we set the incubation time distribution to a Gamma distribution $\Gamma(n,T_\mathrm{inc}/n)$ (see Eq.~(\ref{gamma_distribution})), the results coincide with the solution of the SE$n$IR model from Eq.~(\ref{SEnIR}) with the corresponding $n$. This is shown in Fig.~\ref{full_graph_gamma} for $n=2$ and $n=10$. Fig.~\ref{full_graph_obs} also plots $I_\text{max}$, $t(I_\text{max})$ and $R_\infty$ as a function of $n$. 

\FloatBarrier

\FloatBarrier

We will now focus on the case of scale-free networks, which have a degree distribution $p(k)=\alpha k^{-\gamma}$. The incubation times follow a Gamma distribution $\Gamma(n,T_\mathrm{inc}/n)$  as in the mean-field SE$n$IR model from Section~\ref{section:compartmental}. 

\begin{figure}[ht]\centering
 \begin{subfigure}{.325\textwidth} \centering
 \includegraphics[width=\linewidth]{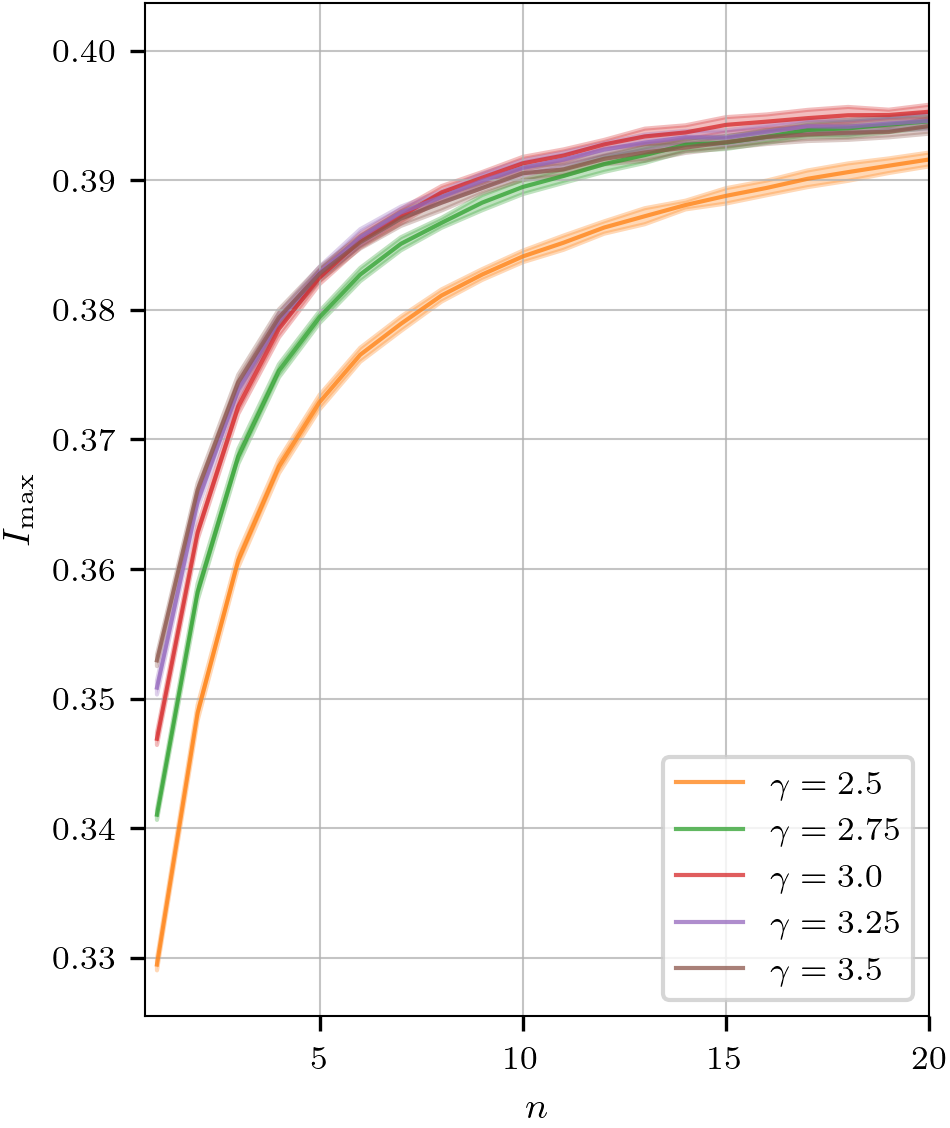}
 \caption{$I_\text{max}$}
 \end{subfigure}
  \begin{subfigure}{.325\textwidth} \centering
 \includegraphics[width=\linewidth]{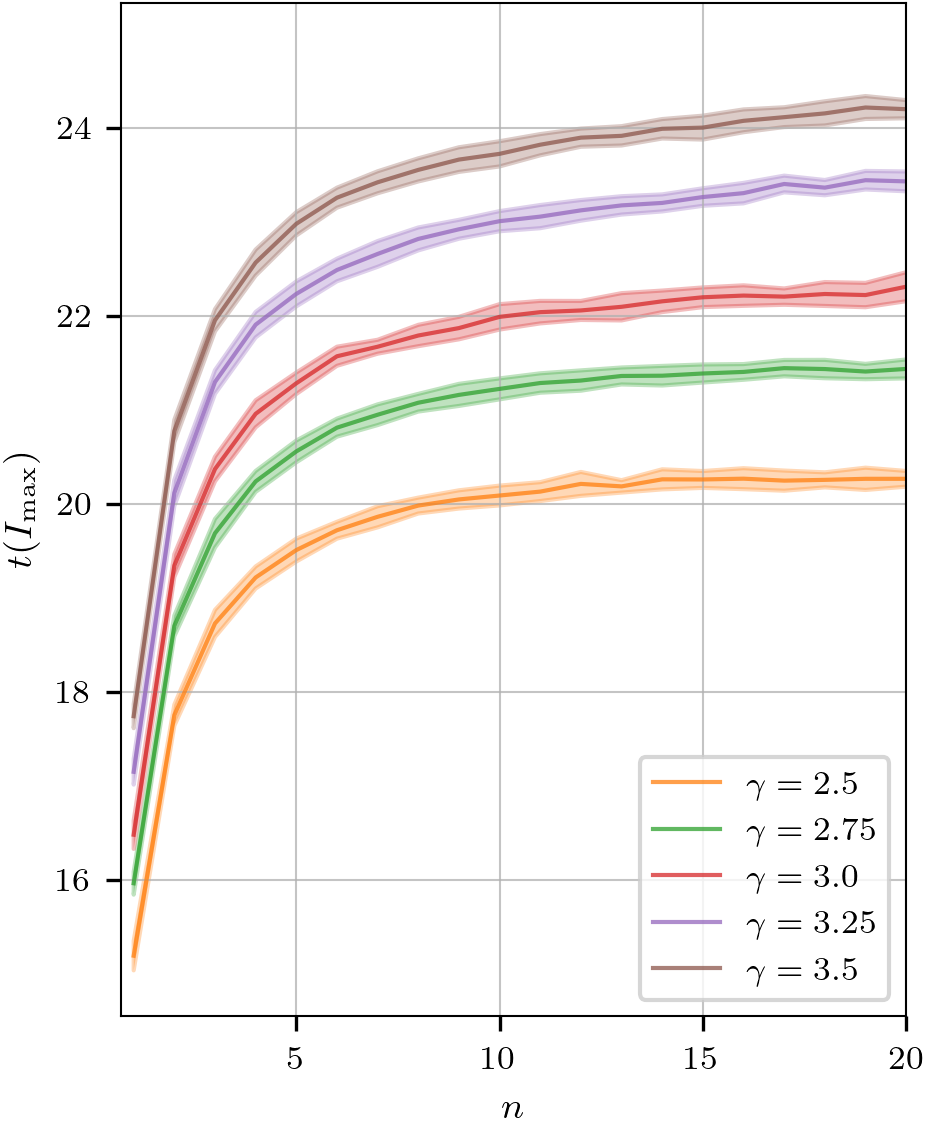}
 \caption{ $t(I_\text{max})$}
 \end{subfigure}
   \begin{subfigure}{.325\textwidth} \centering
 \includegraphics[width=\linewidth]{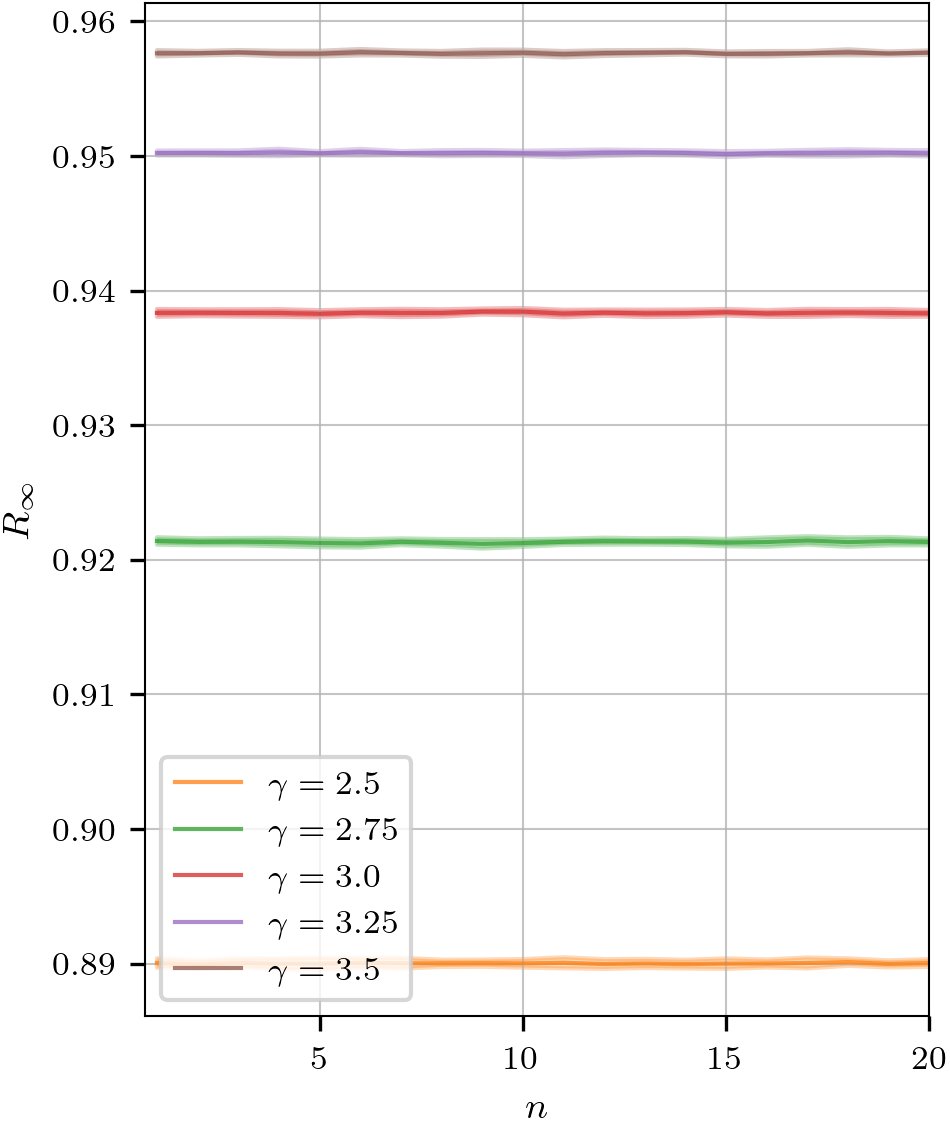}
 \caption{ $R_\infty$}
 \end{subfigure}
 \caption{Relevant quantities as a function of $n$ in networks with degree distribution $p(k)=k^{-\gamma}$ for different values of $\gamma$. }
\label{scale_free}
\end{figure}

Fig.~\ref{scale_free} shows $I_\text{max}$, $t(I_\text{max})$ and $R_\infty$ as a function of $n$ for some values of $\gamma$. The results shown correspond to the average of $30$ realizations of the simulation with each set of parameters.

\FloatBarrier

We can see that for every $\gamma$, the dependency on $n$ is similar to the one shown in Fig.~\ref{full_graph_obs} for fully connected networks. However, in this case, the increase in $I_\text{max}$ and $t(I_\text{max})$ is steeper, which makes it even more important in these cases to be conscious of the incubation time distribution used in the model.

\section{Discussion}
\label{section:discussion}
Throughout the previous sections, we examined the impact of selecting different incubation time distributions in both ODE and agent-based models. In Section~\ref{section:compartmental}, our focus was on the SE$n$IR model, where the incubation times follow a Gamma distribution, as depicted in Eq.~(\ref{gamma_distribution}). There, we observed that the three key metrics —the height of the infection peak, the timing of its occurrence, and the final size of the epidemic— increase as a function of $n$, representing the number of sub-compartments considered for the exposed stage,while holding the doubling rate constant. We propose that this trend arises due to a higher proportion of individuals lingering in the exposed stage for extended durations. Consequently, when the fraction of infected individuals reaches its peak, the prolonged delay within the exposed stage causes the infected curve to surpass its typical peak. This is especially important when noting that most modelers assume $n=1$, and that the exponential distribution is the one that underestimates the aforementioned metrics to the greatest extent.
 
\begin{figure}[ht]\centering
 \begin{subfigure}{.325\textwidth} \centering
 \includegraphics[width=\linewidth]{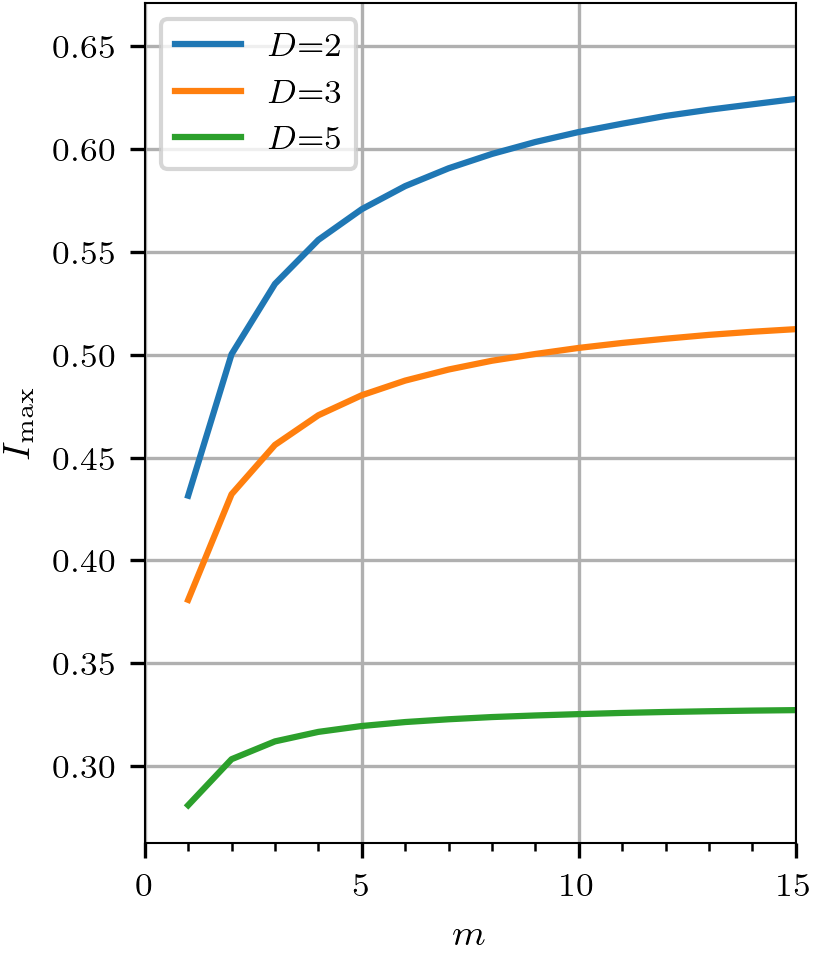}
 \caption{$I_\text{max}$}
 \label{imax_vs_m}
 \end{subfigure}
  \begin{subfigure}{.325\textwidth} \centering
 \includegraphics[width=\linewidth]{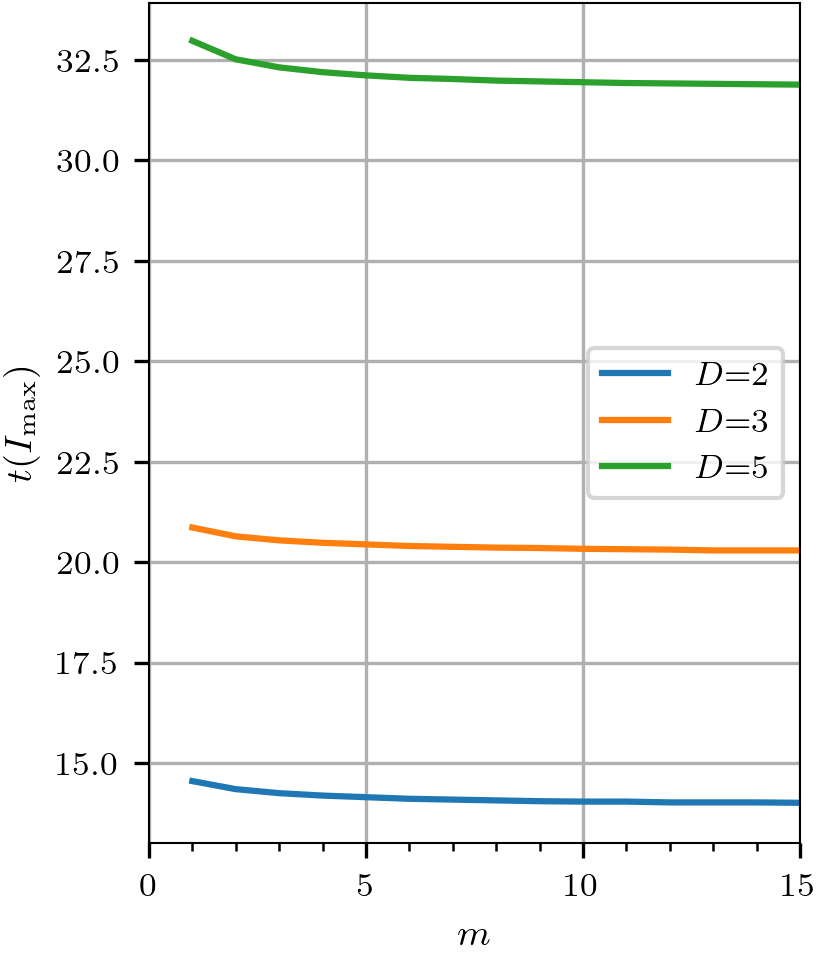}
 \caption{ $t(I_\text{max})$}
 \end{subfigure}
   \begin{subfigure}{.325\textwidth} \centering
 \includegraphics[width=\linewidth]{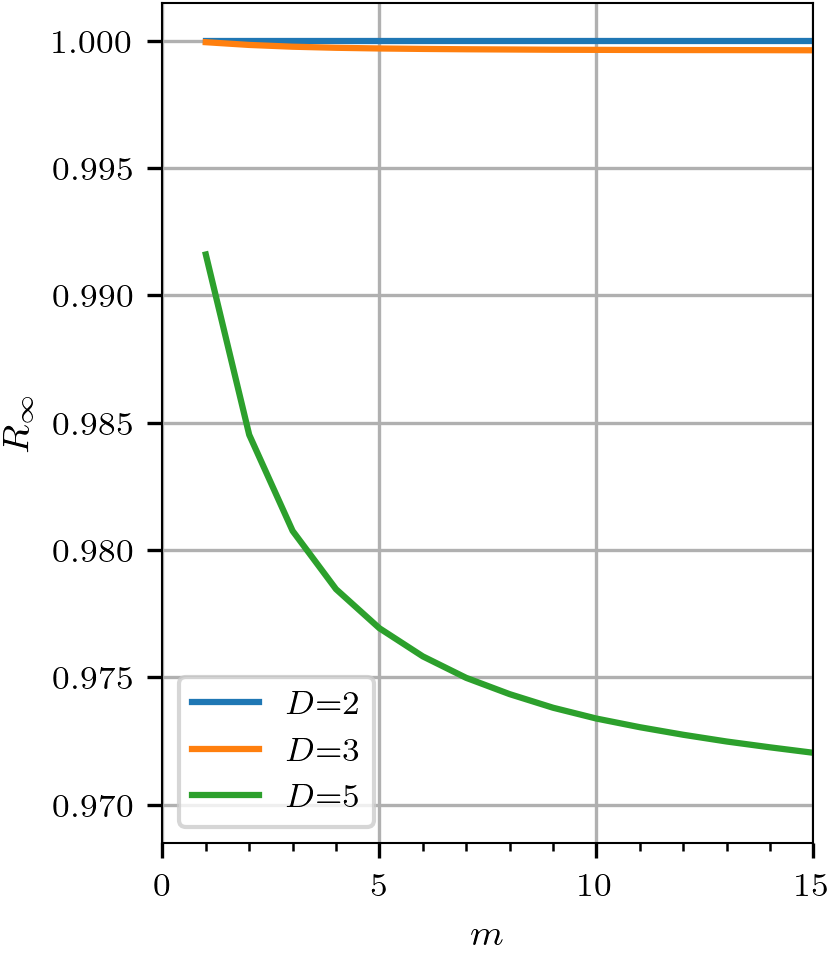}
 \caption{ $R_\infty$}
 \end{subfigure}
 \caption{Relevant quantities as a function of $m$, the number of sub-compartments considered for the infection stage. These results are obtained by including $m$ stages of infection within the classic SEIR ODE model.}
\label{qunatities_vs_m}
\end{figure}

On another front, our study delves into the ramifications of selecting different incubation time distributions within both ODE and agent-based models. Similar questions arise when changing the distribution of the infection times. However, in this investigation, we concentrate solely on the incubation time distribution due to the abundance of field observations and statistical analyses pertaining to the incubation stage. Transitioning from the classic SEIR ODE model to one that incorporates $m$ sub-compartments within the infection stage reveals varying trends in the analyzed metrics as shown in Fig.~\ref{qunatities_vs_m}.  Given the improbability of the distribution being exponential, an $m$ greater than 1 is likely to yield more accurate predictions. This suggests that choosing $m=1$ probably underestimates the peak of infections, as shown in Fig.~\ref{imax_vs_m}.

\FloatBarrier

\section{Conclusion}
\label{section:conclusion}

In this paper, we analyzed how the incubation time distribution influences results in both mean-field and complex network models. In particular, we studied the predictions made by the SE$n$IR mean-field model, which offers a refined version of the SEIR model, and how these predictions vary with different Gamma distributions for the incubation time.

The observed variations in key epidemic predictions, such as the height and timing of infection peaks, underscore the importance of selecting model parameters, namely $n$ and $T_\mathrm{inc}$, to accurately reflect the experimentally measured time distribution during an epidemic outbreak. Our analysis reveals that opting for $n=1$ underestimates the peak of infections, as empirical data suggests that the distribution of incubation times is unlikely to follow an exponential pattern.

Further exploration using agent-based models on complex networks reinforces these findings. The flexibility to customize the time distribution enables us to select the most suitable one in each scenario. 

Comparing exponential and uniform distributions for the incubation time highlights that variations in the distribution can result in significant differences in epidemic predictions. Moreover, extending the analysis to scale-free networks with various degree distributions demonstrates the pronounced impact of the incubation time distribution across network structures.

This research highlights the critical importance of integrating realistic incubation time distributions into epidemic modeling and understanding their implications. When striving for reliable predictions, the choice of incubation time distribution in the model should be carefully considered to inform decisions regarding resource allocation, intervention strategies, and overall epidemic preparedness among professionals relying on mathematical models for decision-making.

\bibliography{refs.bib}
\bibliographystyle{elsarticle-num}

\end{document}